\documentclass[conference]{IEEEtran}
\IEEEoverridecommandlockouts
\usepackage{cite}
\usepackage{amsmath,amssymb,amsfonts}
\usepackage{graphicx}
\usepackage{textcomp}
\usepackage{xcolor}

\usepackage{soul}

\usepackage{algorithm}
\usepackage{algpseudocode}

\usepackage[hidelinks]{hyperref}

\usepackage{amsthm} 
\usepackage{mathtools}

\newtheoremstyle{normalstyle}  % Name of the style
  {}                           % Space above
  {}                           % Space below
  {\normalfont}                % Body font (non-italic)
  {}                           % Indent amount
  {\bfseries}                  % Theorem head font (bold)
  {.}                          % Punctuation after theorem head
  { }                          % Space after theorem head
  {}                           % Theorem head spec

\theoremstyle{normalstyle}
\newtheorem{Definition}{Definition}

\newtheorem{Theorem}{Theorem}

\newtheorem{Remark}{Remark}

\newtheorem{Proposition}{Proposition}
\newtheorem{Lemma}{Lemma}

\def\BibTeX{{\rm B\kern-.05em{\sc i\kern-.025em b}\kern-.08em
    T\kern-.1667em\lower.7ex\hbox{E}\kern-.125emX}}
    
\begin{document}

% Define the notice at the top left of the first page
\IEEEpubid{
\begin{minipage}[t]{\columnwidth}
\begin{flushleft}
\footnotesize\color{red}
This paper has been accepted for presentation at the 64th Conference in Decision and Control (CDC 2025).
\end{flushleft}
\end{minipage}
\hspace{\columnsep}\makebox[\columnwidth]{}
\IEEEpubidadjcol
}

\title{%Less Conservative controlled-invariant Safe Set Synthesis using Control Barrier Certificates //
Minimally Conservative Controlled-Invariant Set Synthesis Using Control Barrier Certificates}

\author{Naeim Ebrahimi Toulkani, and Reza Ghabcheloo
\thanks{Naeim Ebrahimi Toulkani and Reza Ghabcheloo are with the Faculty of Engineering and Natural Sciences, Tampere University,
Korkeakoulunkatu 7 Kampusareena, 33720 Tampere, Finland (e-mail: Naeim.EbrahimiToulkani@tuni.fi; Reza.Ghabcheloo@tuni.fi). }
}

\maketitle

\begin{abstract}
Finding a controlled-invariant set for a system with state and control constraints is crucial for safety-critical applications. However, existing methods often produce overly conservative solutions.  
This paper presents a method for generating controlled-invariant (safe) sets for nonlinear polynomial control-affine systems using Control Barrier Certificates (CBCs). We formulate CBC conditions as Sum-of-Squares (SOS) constraints and solve them via an SOS Program (SOSP).  
First, we generalize existing SOSPs for CBC synthesis to handle environments with complex unsafe state representations. Then, we propose an iterative algorithm that progressively enlarges the safe set constructed by the synthesized CBCs by maximizing boundary expansion at each iteration. We theoretically prove that our method guarantees strict safe set expansion at every step.  
Finally, we validate our approach with numerical simulations in 2D and 3D for single-input and multi-input systems. Empirical results show that the safe set generated by our method covers in most part a larger portion of the state space compared to two state-of-the-art techniques.  

\end{abstract}

%%%%%%%%%%%%%%%%%%
%%%%%%%%%%%%%%%%%%
\section{Introduction}
Safety-critical control systems are extensively used in various fields, including adaptive cruise control \cite{ames2014control}, aerospace \cite{so2024train}, and robotics \cite{toulkani2022reactive}. These systems require rapid and reliable control methods that ensure both safety and optimality. Barrier Certificates (BCs) have been proposed \cite{prajna2004safety} as a means to guarantee safety by identifying invariant subsets within a system’s allowable states. While BCs have been successfully applied to nonlinear autonomous systems, both deterministic and stochastic, there has been limited work extending this method to control systems, i.e. utilizing Control Barrier Certificates (CBCs) with input constraints.

An alternative approach to ensuring safe control in nonlinear systems is the use of Control Barrier Functions (CBFs). They were first proposed by \cite{wieland2007constructive} and later extended to integrate with performance controllers in a Quadratic Program (QP) framework \cite{ames2016control}. Although CBFs can guarantee safety, finding a valid CBF for a nonlinear control system with input constraints is nontrivial. When a CBF is chosen heuristically, the resulting QP may become infeasible during runtime, making the approach unreliable in practice \cite{zhao2023convex}. %This means that the zero superlevel set of a heuristically chosen CBF is not controlled-invariant.

To address infeasibility in practice, Sum-Of-Squares (SOS) decomposition and semidefinite programming \cite{parrilo2000structured} have been proposed for synthesizing CBFs while considering input constraints \cite{zhao2023convex, dai2023convex}. 
In the CBF framework, a class-$\mathcal{K}$\footnote{Recall function $\alpha(.)$ is a class-$\mathcal{K}$ function if it is monotonically increasing and satisfies $\alpha(0)=0$.} function extends the invariance condition from only the boundary of the controlled-invariant set to the entire set. Hence, constructing CBFs with Sum-Of-Squares Programs (SOSPs) requires defining a class-$\mathcal{K}$ function beforehand.  
Consequently, safe sets generated by SOSPs with CBF conditions depend on the chosen class-$\mathcal{K}$ function and tend to be conservative.

% On the contrary, constructing a controlled-invariant set using SOSPs with CBC conditions (original set invariance conditions) imposes constraints only on the boundary of the controlled-invariant set and depends solely on the problem parameters, eliminating the need to define a class-$\mathcal{K}$ function. Therefore, 
% safe sets constructed by CBC conditions can be
% less conservative and yield larger controlled-invariant sets.  

% Therefore, we propose SOSPs to synthesize CBCs and progressively enlarge the safe set, constructed by the synthesized CBCs, in an iterative manner. Unlike the state-of-the-art methods, our enlargement procedure does not rely on a predefined shape (such as inner ellipsoids), and predefined class-$\mathcal{K}$ functions. Therefore, our method results in less conservative controlled-invariant sets.

% To address infeasibility in practice, Sum-Of-Squares (SOS) decomposition and semidefinite programming \cite{parrilo2000structured} have been proposed for synthesizing CBFs while considering input constraints \cite{zhao2023convex, dai2023convex}. In the CBF framework, a class-$\mathcal{K}$ function\footnote{Recall that a function $\alpha(.)$ is class-$\mathcal{K}$ if it is monotonically increasing and satisfies $\alpha(0)=0$.} extends invariance from the boundary to the entire controlled-invariant set. Thus, constructing CBFs with Sum-Of-Squares programs (SOSPs) requires defining a class-$\mathcal{K}$ function beforehand, making the resulting safe sets conservative and dependent on the chosen function.

In contrast, constructing a controlled-invariant set using SOSPs with CBC conditions imposes constraints only on the boundary and depends solely on the problem parameters, eliminating the need for a class-$\mathcal{K}$ function. Therefore, controlled-invariant sets constructed with CBC conditions are less conservative and result in larger safe sets.

We propose SOSPs to synthesize CBCs and iteratively enlarge the safe set constructed by these CBCs. Unlike state-of-the-art methods, our procedure does not rely on predefined shapes (e.g., inner ellipsoids) or class-$\mathcal{K}$ functions, leading to less conservative controlled-invariant sets.
%In addition, in the implementation of our CBCs to safely control the system within the QP framework, any arbitrary class-$\mathcal{K}$ function can be applied to relax the safety condition inside the safe set while ensuring that the safety constraint remains active on the boundary of the controlled-invariant set.
% Therefore, we propose an SOSP that utilizes CBC constraints to synthesize controlled-invariant sets. Additionally, we introduce an iterative algorithm that progressively enlarges the controlled-invariant set generated by the synthesized CBC.
% To implement the constructed CBCs to safely control the system within the QP framework, any arbitrary class-$\mathcal{K}$ function can be applied to relax the safety condition inside the safe set while ensuring that the safety constraint remains active on the boundary of the controlled-invariant set.
To summarise, this paper makes the following contributions:
% \textit{Contributions}:  
% Our main contributions are as follows.  
\begin{itemize}  
\item We propose an SOSP for synthesizing safe sets for polynomial control-affine systems subject to input constraints and a generalized representation of unsafe regions, utilizing CBC conditions. This method can lead to the construction of larger safe sets compared to approaches based on CBF conditions.

%\item We propose a novel SOSP to synthesize CBCs for polynomial control-affine systems with input constraints considering a general format for the set of unsafe states.

%\item We propose an iterative algorithm that progressively enlarges controlled-invariant sets, which does not rely on a predefined shape (such as inner ellipsoids), and predefined class-$\mathcal{K}$ functions, allowing us to construct safe sets with arbitrary shapes that can lead to controlled-invariant sets, covering a larger portion of the set of allowable states.

% \item We propose an iterative algorithm to enlarge safe sets without relying on predefined shapes (e.g., inner ellipsoids) or predefined class-$\mathcal{K}$ functions. This flexibility allows the construction of safe sets with arbitrary shapes, covering a larger portion of the set of allowable states.

\item We propose an iterative algorithm to enlarge safe sets without relying on predefined shapes (e.g., inner ellipsoids) or class-$\mathcal{K}$ functions, enabling the construction of safe sets that cover a larger portion of the allowable states.

\item We demonstrate the efficacy of our approach on single-input 2D and multiple-input 3D dynamical systems. We show that our method outperforms two state-of-the-art methods by generating larger controlled-invariant sets.
\end{itemize}

%%%%%%%%%%%%%%%%%%%%%%%%%%%%%%%%%%%%%%%
\subsection{Related Works}
In general, three main approaches exist for safety certificate synthesis in the literature:  
a) Hamilton-Jacobi (HJ) reachability analysis,  
b) learning-based methods, and  
c) SOSP-based methods.  
Hamilton-Jacobi reachability methods can compute Backward Reachable Sets (BRS) \cite{bansal2017hamilton}, but their time complexity grows exponentially with system size \cite{ganai2024hamilton}, making them impractical for high-dimensional systems.  
Learning-based methods have also been explored for safe controller synthesis \cite{dawson2023safe}. However, these approaches often require long training times and, in most cases, lack formal safety guarantees.  
By comparison, SOSP-based methods offer a more practical alternative \cite{prajna2004safety}, providing solutions in polynomial time while enabling formal safety certification, even in the presence of input constraints.

Most of the existing literature on SOSP applications focuses on synthesizing Lyapunov Functions (LFs) for nonlinear control systems \cite{prajna2004nonlinear, tan2004searching}.  
Recently, Dai et al. proposed an SOSP that uses the contrapositives of the CBF conditions to synthesize safe sets \cite{dai2023convex}. They then maximize the largest inner ellipsoid to enlarge the safe set. However, restricting the enlargement to ellipsoids limits the method’s applicability, particularly in problems where the set of unsafe states has a complex structure, such as those with multiple obstacles.  
Additionally, their approach relies on a predefined class-$\mathcal{K}$ function in the enlargement procedure, which restricts the expansion of the controlled-invariant set and introduces unnecessary conservativeness.

In \cite{zhao2023convex}, the authors proposed a novel algorithm for synthesizing CBFs. They first construct an initial CBF without requiring an initial guess (warm start) and then iteratively enlarge the controlled-invariant set. However, their enlargement procedure
requires defining a class-$\mathcal{K}$ function as a hyperparameter for the algorithm.  

Wang et al. \cite{wang2023safety} introduced an SOSP for synthesizing CBCs in environments where the set of allowable states is defined by a single polynomial in a semialgebraic form. They also formulated an iterative procedure to enlarge the controlled-invariant set.
However, their method is not well suited for cluttered environments (a complex combination of unsafe sets) and their enlargement condition tend to be conservative as the proposed condition for enlargement, searches only for SOS polynomials to satisfy the condition, whereas, we will show that it is possible to reformulate enlargement condition in a way that it is possible to search in the set of all polynomials.
%In this work, we address both of these limitations.  

%In this work, we formulate a safe set synthesis approach based on the barrier certificate concept presented in \cite{prajna2004safety} and propose a new enlargement method capable of handling problems where the set of unsafe states has a complex structure.  

% The rest of the paper is organized as follows. Section \ref{Preliminaries} provides the theoretical background and necessary definitions.  
% In Section \ref{Synthesis}, we propose an SOSP for synthesizing CBCs and introduce an algorithm that progressively enlarges the safe set.  
% Section \ref{Results} presents simulation results on sample systems and analyzes the findings.  
% Finally, Section \ref{Conclusion} summarizes the conclusions of the paper. 

% \textit{Notations}: Sets are denoted by capital calligraphic letters (e.g., $\mathcal{A}$), with boundaries indicated by a partial sign (e.g., $\partial\mathcal{A}$). Matrices are represented by capital letters (e.g., $A$), and polynomials by lowercase letters with their arguments (e.g., $a(\boldsymbol{x})$). $\mathbb{R}$ and $\mathbb{R}^n$ denote the sets of real numbers and $n$-dimensional real vectors, respectively, while $\mathbb{R}[\boldsymbol{x}]$ represents the set of polynomials with real coefficients.

\textit{Notations}: We denote all sets with capital calligraphic letters, such as $\mathcal{A}$; 
all matrices with capital letters, such as $A$;
and all polynomials with lowercase letters with their arguments, such as $a(\boldsymbol{x})$.
In addition, $\mathbb{R}$ and $\mathbb{R}^n$ denote sets of real numbers and $n$-dimensional real vectors, respectively.
Finally, $\mathbb{R}[\boldsymbol{x}]$ denotes the set of polynomials with real coefficients. 
%%%%%%%%%%%%%%%%%%%%%%%%%%%%%%%%%%%%%%%%%% 
%%%%%%%%%%%%%%%%%%%%%%%%%%%%%%%%%%%%%%%%%%
\section{Preliminaries and Problem Formulation}
\label{Preliminaries}
%In this section, we first introduce definitions that will be used throughout the paper. We then formulate the problem and present the necessary mathematical tools that will be utilized in Section~\ref{Synthesis}.

In this section, we introduce definitions and the necessary mathematical tools that will be used throughout the paper. We will also formulate our problem.

\begin{Definition}
    A polynomial $p(\boldsymbol{x})$ is said to be a Sum-Of-Squares (SOS) polynomial if and only if there exist polynomials $q_i(\boldsymbol{x}) \ (i = 1, \ldots, n_q)$ such that $p(\boldsymbol{x}) =\sum_{i=1}^{n_q} q_i(\boldsymbol{x})^2$.
    The set of SOS polynomials in the variables $\boldsymbol{x}$ is denoted by $\Sigma[\boldsymbol{x}]$.
\end{Definition}

\begin{Definition}
    A set $\mathcal{X}$ is semialgebraic if it can be represented using polynomial equality and inequality constraints. If it is defined solely by equality constraints, the set is algebraic.
\end{Definition}

The problem of determining whether a polynomial can be decomposed into an SOS polynomial is equivalent to checking the existence of a quadratic representation $p(\boldsymbol{x}) = Z^T(\boldsymbol{x})QZ(\boldsymbol{x})$ where $Q$ is a Positive Semi-Definite (PSD) matrix and $Z(\boldsymbol{x})$ is a vector of monomials \cite{parrilo2000structured}.  

\begin{Remark}
    If a polynomial $p(\boldsymbol{x})$ is an SOS polynomial, then it is necessarily non-negative for all $\boldsymbol{x}$ in its domain. Furthermore, the expression $\sum_i q_i(\boldsymbol{x})^2$ represents the SOS decomposition of $p(\boldsymbol{x})$.
    %However, not every non-negative polynomial can be decomposed as an SOS polynomial \cite{parrilo2000structured}.
\end{Remark}

\begin{Definition}
    The optimization problem that determines whether a polynomial $p(\boldsymbol{x})$ can be expressed as $\sum_i q_i(\boldsymbol{x})^2$ %by searching for a PSD matrix $Q$ 
    is called an SOSP.  
    An SOSP 
    %can be reformulated as a Semi Definite Program (SDP), which 
    is a convex program.
\end{Definition}

\begin{Lemma}[S-Procedure \cite{parrilo2000structured}]
    A sufficient condition to establish that $p(\boldsymbol{x}) \geq 0$ on the semialgebraic set $\mathcal{K} = \{\boldsymbol{x} \mid a_1(\boldsymbol{x}) \geq 0, \dots, a_q(\boldsymbol{x}) \geq 0\}$ is the existence of $\sigma_i(\boldsymbol{x}) \in \Sigma[\boldsymbol{x}]$ for $i = 1, \dots, q$, such that  
    \vspace{-0.2 cm}
    \begin{equation*}
        p(\boldsymbol{x}) - \sum_{i=1}^{q} \sigma_i(\boldsymbol{x}) a_i(\boldsymbol{x}) \in \Sigma[\boldsymbol{x}]. \tag{$\ast$}
    \end{equation*}
    \label{S_Procedure_Lemma}
\end{Lemma}
\vspace{-0.7 cm}
% \begin{Remark}
%     When the set $\mathcal{K}$ is algebraic, i.e., $\mathcal{K} = \{\boldsymbol{x} \mid b_1(\boldsymbol{x}) = 0, \dots, b_q(\boldsymbol{x}) = 0\}$,
%     %where $q$ is a finite positive integer, 
%     the multipliers $\sigma_i(\boldsymbol{x})$ as defined in Lemma \ref{S_Procedure_Lemma}, can be any polynomial, not necessarily an SOS polynomial.  
% \end{Remark}

\begin{Remark}
    In Lemma \ref{S_Procedure_Lemma}, if set $\mathcal{K}$ is algebraic, i.e., $\mathcal{K} = \{\boldsymbol{x} \mid a_1(\boldsymbol{x}) = 0, \dots, a_q(\boldsymbol{x}) = 0\}$, then condition \textnormal{(}{$\ast$}\textnormal{)} is a 
    sufficient for $p(\boldsymbol{x}) \geq 0$ on $\mathcal{K}$  if there exist $\sigma_i(\boldsymbol{x}) \in \mathbb{R}[\boldsymbol{x}]$ for $i = 1, \dots, q$.
\end{Remark}

Finally, the notion of controlled-invariant sets can be defined as follows:

\begin{Definition}
    A set of states $\mathcal{S}$ is controlled-invariant for a control system if, for all initial states in $\mathcal{S}$, there exists a control action that ensures the system trajectory remains within $\mathcal{S}$ for all future time \cite{blanchini1999set}.
\end{Definition}

%%%%%%%%%%%%%%%%%%%%%%%%
\subsection{Problem Setting}
Consider a nonlinear control-affine system:
\begin{equation}
\label{eq:affine_sys}
\dot{\boldsymbol{x}} = f(\boldsymbol{x}) + g(\boldsymbol{x})\boldsymbol{u},
\end{equation}
where $\boldsymbol{x}(t) \in \mathcal{X} \subseteq \mathbb{R}^{n_x}$ and $\boldsymbol{u}(t) \in \mathcal{U} \subseteq \mathbb{R}^{n_u}$. The functions $f: \mathbb{R}^{n_x} \to \mathbb{R}^{n_x}$ and $g: \mathbb{R}^{n_x} \to \mathbb{R}^{n_x\times n_u}$ are polynomial functions. The set of input constraints for system \eqref{eq:affine_sys} is defined as:
\begin{equation}
\label{eq:input_constraint}
\mathcal{U} = \{\boldsymbol{u} \in \mathbb{R}^{n_u} \mid A_u\boldsymbol{u} + c_u \geq 0\},
\end{equation}
where $A_u \in \mathbb{R}^{\tau \times {n_u}}$ and $c_u \in \mathbb{R}^\tau$.

\begin{Remark}
    The input constraint set $\mathcal{U}$ can be state-dependent if $A_u = A_u(\boldsymbol{x})$ and $c_u = c_u(\boldsymbol{x})$. While our derivations remain valid for state-dependent input constraints, we assume constant $A_u$ and $c_u$ for ease of presentation.
\end{Remark}

Our problem is to construct a safe set for system \eqref{eq:affine_sys} in a set of allowable states. Next, we define notions of 
safe, unsafe, and allowable state sets.% for system \eqref{eq:affine_sys} are defined as follows:

\begin{Definition}
    The set of \underline{unsafe states}, $\mathcal{X}_u$, is given by
    \begin{equation}
    \label{eq:unsafe_set}
        \mathcal{X}_u = \mathcal{X}_{u_1} \cup \dots \cup \mathcal{X}_{u_n} = \bigcup_{i=1}^{n} \mathcal{X}_{u_i},
    \end{equation}
    where $\mathcal{X}_{u_i}$ is a semialgebraic set with $m_i$ inequalities:
    \begin{equation}
    \label{eq:unsafe_set2}
    \mathcal{X}_{u_i} = \{\boldsymbol{x} \in \mathbb{R}^{n_x} \mid s_{i,1}(\boldsymbol{x}) < 0,\dots,s_{i,m_i}(\boldsymbol{x}) < 0\}.
    \end{equation}  
\end{Definition}
\begin{figure}
    \centering
    \includegraphics[width=0.75\columnwidth, trim= 0.2cm 8.3cm 20cm 0.2cm, clip]{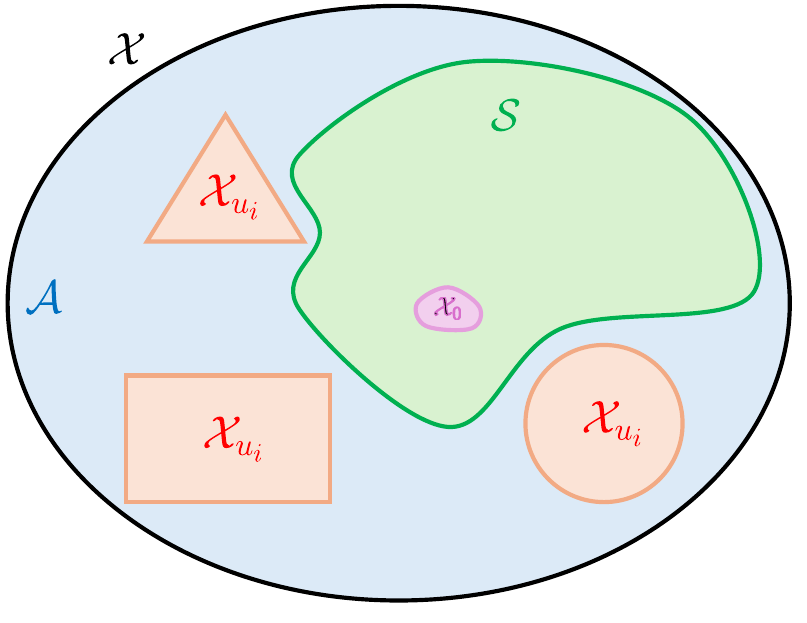}
    \caption{Graphical illustration of the domain set $\mathcal{X}$ (black) and the sets of safe $\mathcal{S}$ (green), unsafe $\mathcal{X}_u$ (red), allowable $\mathcal{A}$ (blue), and initial states $\mathcal{X}_0$ (purple).}
    \label{fig:sets}
    \vspace{-0.3 cm}
\end{figure}

\begin{Definition}
    The set of \underline{allowable states}, denoted as $\mathcal{A}$, is the complement of the unsafe set, i.e., $\mathcal{A} = \Bar{\mathcal{X}}_u$.
\end{Definition}

% \begin{Definition}
%     A set $\mathcal{S} \subseteq \mathcal{A}$ is called a safe set for control system \eqref{eq:affine_sys} if it is controlled-invariant.
%     \textcolor{red}{Notions of Safe set and Controlled-invariant set are used interchangeably throughout the paper.} 
% \end{Definition}

\begin{Definition}
    A set $\mathcal{S} \subseteq \mathcal{A}$ is a safe set for control system \eqref{eq:affine_sys} if it is controlled-invariant. The terms \underline{safe set} and \underline{controlled-invariant set} are used interchangeably in this paper.  
\end{Definition}

Figure~\ref{fig:sets} provides a graphical illustration of the sets, which will be used throughout the paper.
Now, we will continue with defining the notion of Control Barrier Certificates (CBCs).

% \begin{Definition} [CBCs \cite{prajna2004safety}]
%      Let system \eqref{eq:affine_sys}, set of unsafe states $\mathcal{X}_u$, set of input constraints $\mathcal{U}$, and the set of initial states $\mathcal{X}_0$ be given. Then a function $b(\boldsymbol{x}): \mathcal{X} \to \mathbb{R}$ is called a Control Barrier Certificate for system \eqref{eq:affine_sys}, if it satisfies:% the following conditions:
% \begin{subequations}
% \label{eq:CBC_conds}
% \begin{equation}
%     \label{eq:BC_cond_1}
%         b(\boldsymbol{x}) < 0 \quad \forall \boldsymbol{x} \in \mathcal{X}_u,
%     \end{equation}
%     \begin{equation}
%     \label{eq:BC_cond_2}
%         b(\boldsymbol{x}) \ge 0 \quad \forall \boldsymbol{x} \in \mathcal{X}_0,
%     \end{equation}
%     \begin{equation}
%     \begin{multlined}
%     \label{eq:BC_cond_3}
%         \forall \boldsymbol{x} \textnormal{ s.t. } b(\boldsymbol{x}) = 0, \ \exists \boldsymbol{u} \in \mathcal{U} \textnormal{ such that:} \\ 
%         \frac{\partial b(\boldsymbol{x})}{\partial \boldsymbol{x}}(f(\boldsymbol{x}) + g(\boldsymbol{x}) \boldsymbol{u}) \ge 0.\quad
%     \end{multlined}
%     \end{equation}
%     \label{eq:CBC_definition}
% \end{subequations}
% \label{CBC_definition}
% \end{Definition}

\begin{Definition} [CBCs \cite{prajna2004safety}]
    Given a system \eqref{eq:affine_sys}, the unsafe set $\mathcal{X}_u$, input constraints $\mathcal{U}$, and initial states $\mathcal{X}_0$, a function $b: \mathcal{X} \to \mathbb{R}$ is a Control Barrier Certificate (CBC) for the system \eqref{eq:affine_sys} if:
    \vspace{-0.2 cm}
\begin{subequations}
\label{eq:CBC_conds}
    \begin{equation}
        b(\boldsymbol{x}) < 0, \quad \forall \boldsymbol{x} \in \mathcal{X}_u, \label{eq:BC_cond_1}
    \end{equation}
    \begin{equation}
        b(\boldsymbol{x}) \ge 0, \quad \forall \boldsymbol{x} \in \mathcal{X}_0, \label{eq:BC_cond_2}
    \end{equation}
    \begin{equation}
        \forall \boldsymbol{x} \text{ with } b(\boldsymbol{x}) = 0, \ \exists \boldsymbol{u} \in \mathcal{U} \text{ s.t. } \frac{\partial b}{\partial \boldsymbol{x}}(f(\boldsymbol{x}) + g(\boldsymbol{x}) \boldsymbol{u}) \ge 0. \label{eq:BC_cond_3}
    \end{equation}
\end{subequations}
\label{CBC_definition}
\end{Definition}
\vspace{-0.6 cm}

Ensuring safety in dynamical systems requires guaranteeing that system trajectories never enter unsafe regions while respecting control limitations. Control Barrier Certificates (CBCs) provide a framework for designing such control strategies. We begin by recalling a fundamental result from \cite{prajna2004safety}.

\begin{Theorem}[Theorem 1 of \cite{prajna2004safety}]
    If there exists a function $b(\boldsymbol{x})$ that satisfies conditions \eqref{eq:CBC_conds}, then the system \eqref{eq:affine_sys} is guaranteed to be safe. That is, no trajectory starting in $\mathcal{X}_0$ reaches an unsafe state in $\mathcal{X}_u$.
\label{Jadbabaie}
\end{Theorem}

%In the next section, we will explain how CBCs can guarantee system safety and we will develop an SOSP to synthesize CBCs for arbitrary polynomial control-affine systems.
%%%%%%%%%%%%%%%%%%%%%%%
%%%%%%%%%%%%%%%%%%%%%%%
%%%%%%%%%%%%%%%%%%%%%%%
\section{Developing SOSP for CBC Synthesis Under Input Constraints}
\label{Synthesis}

% Ensuring safety in dynamical systems requires guaranteeing that system trajectories never enter unsafe regions while respecting control limitations. Control Barrier Certificates (CBCs) provide a framework for designing such control strategies. We begin by recalling a fundamental result from \cite{prajna2004safety}.

% \begin{Theorem}[Theorem 1 of \cite{prajna2004safety}]
%     If there exists a function $b(\boldsymbol{x})$ that satisfies conditions \eqref{eq:CBC_conds}, then the system \eqref{eq:affine_sys} is guaranteed to be safe. That is, no trajectory starting in $\mathcal{X}_0$ reaches an unsafe state in $\mathcal{X}_u$.
% \label{Jadbabaie}
% \end{Theorem}

In the following subsections, we first propose an SOSP formulation for synthesizing CBCs. Then, we introduce an algorithm that iteratively enlarges the CBC-based safe sets for a given dynamical system.

\subsection{Developing an SOSP for CBC Synthesis}
% Synthesizing CBCs 
% can be efficiently addressed using SOS programming. 
% In SOSPs, the positivity constraint on a set is relaxed into a feasibility problem of decomposing a polynomial into a sum of squares, as described by Lemma \ref{S_Procedure_Lemma} (S-Procedure).  
% In the following theorem, we generalize the SOSP from \cite{wang2023safety} for CBC synthesis, allowing us to handle problems with a more complex set of unsafe state representations fits it better for use in cluttered environments.  
Synthesizing CBCs can be efficiently addressed using SOS programming. 
In SOSPs, enforcing positivity over a set is reformulated as a feasibility problem of decomposing a polynomial into a sum of squares, as described in Lemma \ref{S_Procedure_Lemma} (S-Procedure).  
In the following theorem, we generalize the SOSP from \cite{wang2023safety} for CBC synthesis, enabling it to handle more complex representations of unsafe states, making it better suited for cluttered environments.

\begin{Theorem}
Let system \eqref{eq:affine_sys}, a control-admissible set $\mathcal{U}$ as defined in \eqref{eq:input_constraint}, an unsafe set $\mathcal{X}_u$ as defined in \eqref{eq:unsafe_set}, and an initial state set $\mathcal{X}_0$ defined as the zero superlevel set of a polynomial function $\phi(\boldsymbol{x})$, i.e., $\mathcal{X}_0=\{\boldsymbol{x} \mid \phi(\boldsymbol{x})\ge0\}$, be given.
If 
\begin{subequations}
\label{eq:CBC_synth}
    \begin{equation*}
    \begin{multlined}
         \textnormal{Find } b(\boldsymbol{x}), u(\boldsymbol{x}), \lambda_1(\boldsymbol{x}), \lambda_2(\boldsymbol{x}) \in \mathbb{R}[\boldsymbol{x}], \\ 
         \textnormal{and } \sigma_{i,j}(\boldsymbol{x}), \sigma_2(\boldsymbol{x}) \in \Sigma[\boldsymbol{x}], \textnormal{ s.t. } \qquad \quad 
    \end{multlined}
    \end{equation*}
    \begin{equation}
    \label{eq:theo_a}
      -b(\boldsymbol{x}) + \sum_{j=1}^{m_i}{\sigma_{i,j}(\boldsymbol{x}) s_{i,j}(\boldsymbol{x})} -\epsilon \in \Sigma[\boldsymbol{x}], \quad i = 1, \ldots, n,
    \end{equation}
    \begin{equation}
    \label{eq:theo_a2}
     b(\boldsymbol{x}) - \sigma_2(\boldsymbol{x}) \phi(\boldsymbol{x}) \in \Sigma[\boldsymbol{x}],
    \end{equation}
    \begin{equation}
        \label{eq:theo_c}
     A_u \boldsymbol{u}(\boldsymbol{x}) + c_u - \lambda_1(\boldsymbol{x}) b(\boldsymbol{x}) \in \Sigma[\boldsymbol{x}],
    \end{equation}
    \begin{equation}
    \label{eq:theo_b}
     \frac{\partial b(\boldsymbol{x})}{\partial \boldsymbol{x}}(f(\boldsymbol{x}) + g(\boldsymbol{x}) \boldsymbol{u}(\boldsymbol{x})) - \lambda_2(\boldsymbol{x}) b(\boldsymbol{x}) \in \Sigma[\boldsymbol{x}].
    \end{equation}
\end{subequations}
is feasible for some $\epsilon>0$, then $b(\boldsymbol{x})$ is a CBC for system \eqref{eq:affine_sys} and set 
% \begin{equation}
%     \mathcal{S}=\{\boldsymbol{x} \mid b(\boldsymbol{x})\ge0\}
% \end{equation}
$\mathcal{S}=\{\boldsymbol{x} \mid b(\boldsymbol{x})\ge0\}$
is a safe set for the system.
\label{CBC_synth}
\end{Theorem}

\begin{proof}
    Let $b^*, u^*, \lambda_1^*, \lambda_2^*, \sigma_{i,j}^*$, and $\sigma_2^*$ be the solutions of SOSP \eqref{eq:CBC_synth}. We show that conditions \eqref{eq:BC_cond_1} to \eqref{eq:BC_cond_3} hold. To prove that $b^*(\boldsymbol{x})<0$ for $\boldsymbol{x} \in \mathcal{X}_u$, assume without loss of generality that $\boldsymbol{x} \in \mathcal{X}_{u_i}$ for some $i \in \{1,\ldots,n\}$. Then,  
    \begin{equation*}
    -b^*(\boldsymbol{x}) + \sum_{j=1}^{m_i} \sigma^*_{i,j}(\boldsymbol{x}) s_{i,j}(\boldsymbol{x}) -\epsilon \in \Sigma[\boldsymbol{x}].
    \end{equation*}
    By Lemma \ref{S_Procedure_Lemma}, it follows that $-b^*(\boldsymbol{x}) - \epsilon\geq0$ for $\boldsymbol{x} \in \mathcal{X}_{u_i}$, which implies $b^*(\boldsymbol{x})<0$ for $\boldsymbol{x} \in \mathcal{X}_{u_i}$. Therefore, condition \eqref{eq:BC_cond_1} holds for $\boldsymbol{x} \in \mathcal{X}_{u_i}$, $i \in \{1,2,\ldots,n\}$. Since $\mathcal{X}_u$ is the union of these sets, condition \eqref{eq:BC_cond_1} holds.  

    To show that condition \eqref{eq:BC_cond_2} is satisfied, note that  
    \begin{equation*}
        b^*(\boldsymbol{x}) - \sigma^*_2(\boldsymbol{x}) \phi(\boldsymbol{x}) \in \Sigma[\boldsymbol{x}].
    \end{equation*}
    By Lemma \ref{S_Procedure_Lemma}, it follows that $b^*(\boldsymbol{x}) \geq 0$ for all $\boldsymbol{x} \in \mathcal{X}_0$.  

    To establish the satisfaction of condition \eqref{eq:BC_cond_3}, we first show that $\boldsymbol{u}^* \in \mathcal{U}$ and then that  
    \begin{equation*}
        \frac{\partial b^*(\boldsymbol{x})}{\partial \boldsymbol{x}}(f(\boldsymbol{x}) + g(\boldsymbol{x})\boldsymbol{u}^*(\boldsymbol{x}))\geq0 \quad \text{for } \boldsymbol{x} \in \{\boldsymbol{x} \mid b^*(\boldsymbol{x})=0\}.
    \end{equation*}
    According to \eqref{eq:theo_c},  
    \begin{equation*}
        A_u \boldsymbol{u}^*(\boldsymbol{x}) + c_u - \lambda_1^*(\boldsymbol{x}) b^*(\boldsymbol{x}) \in \Sigma[\boldsymbol{x}].
    \end{equation*}
    By Lemma \ref{S_Procedure_Lemma}, this implies $A_u \boldsymbol{u}^*(\boldsymbol{x}) + c_u\geq0$, and by \eqref{eq:input_constraint}, we conclude that $\boldsymbol{u}^*(\boldsymbol{x}) \in \mathcal{U}$ for all $\boldsymbol{x} \in \{\boldsymbol{x} \mid b^*(\boldsymbol{x})=0\}$.  

    According to \eqref{eq:theo_b},  
    \begin{equation*}
        \frac{\partial b^*(\boldsymbol{x})}{\partial \boldsymbol{x}}(f(\boldsymbol{x}) + g(\boldsymbol{x}) \boldsymbol{u}^*(\boldsymbol{x})) - \lambda^*_2(\boldsymbol{x}) b^*(\boldsymbol{x}) \in \Sigma[\boldsymbol{x}].
    \end{equation*}
    By Lemma \ref{S_Procedure_Lemma}, we obtain  
    \begin{equation*}
        \frac{\partial b^*(\boldsymbol{x})}{\partial \boldsymbol{x}}(f(\boldsymbol{x}) + g(\boldsymbol{x}) \boldsymbol{u}^*(\boldsymbol{x}))\geq 0 \quad \text{for } \boldsymbol{x} \in \{\boldsymbol{x} \mid b^*(\boldsymbol{x})=0\}.
    \end{equation*}
    Thus, condition \eqref{eq:BC_cond_3} holds and $b^*(\boldsymbol{x})$ is a CBC and $\mathcal{S}=\{\boldsymbol{x} \mid b^*(\boldsymbol{x})\ge0\}$ is a safe set for system \eqref{eq:affine_sys}.
\end{proof}

% \begin{Remark}
%     Note that according to Theorem \ref{Jadbabaie}, set $\mathcal{S}=\{\boldsymbol{x} \mid b(\boldsymbol{x}) \ge 0 \}$, where $b(\boldsymbol{x})$ synthesized using SOSP \eqref{eq:CBC_synth}, is a safe set for the control system \eqref{eq:affine_sys}.
%     %, since there always exists a control input on the boundary of $\mathcal{S}$ that keeps the system trajectories inside $\mathcal{S}$.
% \end{Remark}

% \begin{Remark}
%     The SOSP for CBC synthesis proposed in \cite{wang2023safety} is a special case of our formulation, where the allowable region is a semialgebraic set defined by a single polynomial.
%     \textcolor{red}{Now that I have changed the paragraph before theorem 2, I think can remove this remark.}
% \end{Remark}

The challenge in solving SOSP \eqref{eq:CBC_synth} arises from the bilinear terms $\lambda_1(\boldsymbol{x}) b(\boldsymbol{x})$ in \eqref{eq:theo_c}, and $\frac{\partial b(\boldsymbol{x})}{\partial \boldsymbol{x}}(f(\boldsymbol{x}) + g(\boldsymbol{x}) \boldsymbol{u}(\boldsymbol{x}))$ and $\lambda_2(\boldsymbol{x}) b(\boldsymbol{x})$ in \eqref{eq:theo_b}, which render the problem non-convex. A common approach to addressing such non-convex problems is bilinear alternation \cite{dai2023convex, zhao2023convex}. This method involves fixing a part of the bilinear terms, thereby making the remaining terms linear in the unknowns, which allows the problem to be solved by available solvers. The process then iterates by alternating the unknowns and using the solution from the previous iteration.

Beyond solving the SOSP, our goal is to construct the largest possible safe set $\mathcal{S}$ as defined in Theorem \ref{CBC_synth}. To achieve this, we propose an iterative procedure that expands $\mathcal{S}$ at each step. Before detailing our method, we establish a condition that ensures set enlargement in each iteration. This leads to our first result, presented as follows.  

% Beyond solving the SOSP, our goal is to construct the largest possible safe set $\mathcal{S}=\{\boldsymbol{x} \mid b(\boldsymbol{x}) \geq 0\}$. Hence, we propose an iterative procedure that expands $\mathcal{S}$ at each iteration. Before detailing our iterative method, we need a condition that ensures set enlargement in the iterations.
% % we must establish how to define SOS conditions that ensure the safe set $\mathcal{S}^{(k)}=\{\boldsymbol{x} \mid b^{(k)}(\boldsymbol{x}) \geq 0\}$, produced in iteration $(k)$, is a subset of the safe set $\mathcal{S}^{(k+1)}=\{\boldsymbol{x} \mid b^{(k+1)}(\boldsymbol{x}) \geq 0\}$, produced in iteration $(k+1)$.  
% % To facilitate this expansion within our SOSP, 
% This leads to our first result presented as follows.
% %we present the following proposition.

\begin{Proposition}
\label{proposition1}
    Assume that $b^{(k)}(\boldsymbol{x})$ and $b^{(k+1)}(\boldsymbol{x})$ are two polynomial CBCs as defined in \eqref{eq:CBC_conds}, and $\mathcal{S}^{(k)}=\{\boldsymbol{x} \mid b^{(k)}(\boldsymbol{x}) \geq 0\}$ and $\mathcal{S}^{(k+1)}=\{\boldsymbol{x} \mid b^{(k+1)}(\boldsymbol{x}) \geq 0\}$.
    %that $\mathcal{S}^{(k)}$ and $\mathcal{S}^{(k+1)}$ are their zero superlevel sets, respectively. 
    If the conditions  
    \begin{subequations}
    \label{eq:enlargement}
        \begin{equation}
        \label{eq:enlargement1}
            b^{(k+1)}(\boldsymbol{x}) - \mu(\boldsymbol{x}) b^{(k)}(\boldsymbol{x}) \in \Sigma[\boldsymbol{x}], \quad \mu(\boldsymbol{x}) \in \Sigma[\boldsymbol{x}],
        \end{equation}
        \text{and}
        \begin{equation}
        \label{eq:enlargement2}
            b^{(k+1)}(\boldsymbol{x}) - \lambda(\boldsymbol{x}) b^{(k)}(\boldsymbol{x}) - \gamma \in \Sigma[\boldsymbol{x}], \quad \lambda(\boldsymbol{x}) \in \mathbb{R}[\boldsymbol{x}], \ \gamma > 0,
        \end{equation}
    \end{subequations}
    hold, then $\mathcal{S}^{(k)} \subset \mathcal{S}^{(k+1)}$.  
    In addition, the value of $b^{(k+1)}(\boldsymbol{x})$ at the boundary of $\mathcal{S}^{(k)}$ is greater than or equal to $\gamma$.
\end{Proposition}

\begin{proof}
    Similar to the proof of Theorem \ref{CBC_synth}, and using Lemma \ref{S_Procedure_Lemma}, the satisfaction of condition \eqref{eq:enlargement1} implies that $b^{(k+1)}(\boldsymbol{x}) > 0$ for all $\boldsymbol{x} \in \mathcal{S}^{(k)}$. Hence, $\mathcal{S}^{(k)} \subseteq \mathcal{S}^{(k+1)}$.

    According to Lemma \ref{S_Procedure_Lemma}, the satisfaction of \eqref{eq:enlargement2} implies that $b^{(k+1)}(\boldsymbol{x}) - \gamma \geq 0$, and therefore, $b^{(k+1)}(\boldsymbol{x}) > 0$ on the boundary of $\mathcal{S}^{(k)}$. %i.e., $\partial\mathcal{S}^{(k)} = \{\boldsymbol{x} \mid b^{(k)}(\boldsymbol{x}) = 0\}$.
    Thus, the satisfaction of both conditions \eqref{eq:enlargement1} and \eqref{eq:enlargement2} ensures that $\mathcal{S}^{(k)} \subset \mathcal{S}^{(k+1)}$.
\end{proof}

\begin{figure}
    \centering
    \includegraphics[width=0.75\columnwidth, trim= 0.15cm 4.82cm 24cm 0.2cm, clip]{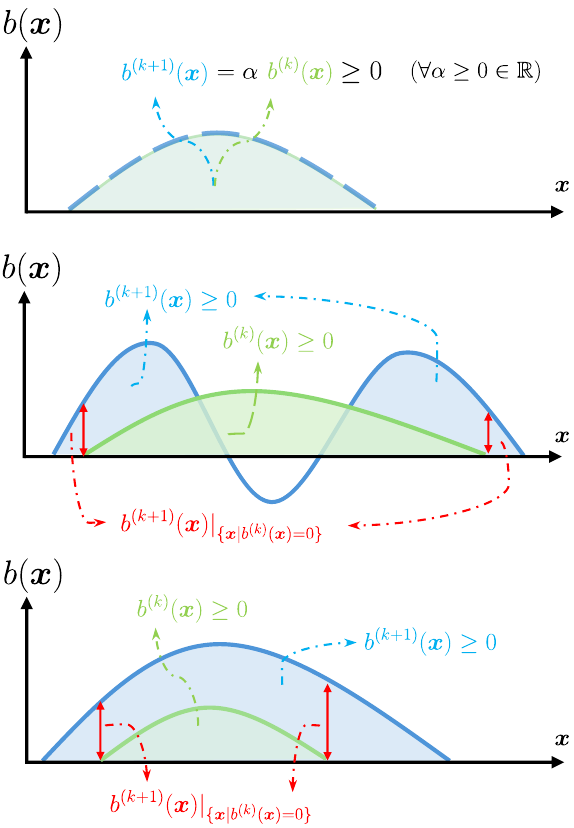}
    \caption{Illustrative 1D example of enlarging safe set from iteration $(k)$ and $(k+1)$, see Proposition \ref{proposition1}.
     The top graph illustrates a case where condition \eqref{eq:enlargement1} alone fails to enlarge the safe set. The middle graph explains why condition \eqref{eq:enlargement1} is insufficient for the enlargement procedure. The bottom graph shows how applying both conditions from Proposition \ref{proposition1} guarantees set enlargement.}
    \label{fig:B_k_and_B_k1}
    \vspace{-0.3 cm}
\end{figure}

Figure~\ref{fig:B_k_and_B_k1} provides further intuition for Proposition \ref{proposition1}, assuming that we are enlarging a CBC for a one-dimensional system, i.e., $b(\boldsymbol{x}):\mathbb{R} \to \mathbb{R}$, between iterations $(k)$ and $(k+1)$.  

It should be noted that using condition \eqref{eq:enlargement1} alone may end up having a trivial answer for the enlargement. Figure~\ref{fig:B_k_and_B_k1} (top graph) indicates one example of such a trivial situation.
Furthermore, condition \eqref{eq:enlargement2} alone cannot guarantee that $\mathcal{S}^{(k)} \subset \mathcal{S}^{(k+1)}$, as demonstrated by the example in Fig.~\ref{fig:B_k_and_B_k1} (middle graph).
Therefore, both conditions are necessary to ensure the enlargement of the set between iterations Fig.~\ref{fig:B_k_and_B_k1} (bottom graph).

\begin{Remark}
    One can suggest to have a condition as
    $b^{(k+1)}(\boldsymbol{x}) - \mu(\boldsymbol{x}) b^{(k)}(\boldsymbol{x}) - \gamma \in \Sigma[\boldsymbol{x}], \quad \mu(\boldsymbol{x}) \in \Sigma[\boldsymbol{x}], \ \gamma > 0,$
    instead of conditions \eqref{eq:enlargement1} and \eqref{eq:enlargement2}. However, satisfying such a condition is much harder than satisfying \eqref{eq:enlargement2}, as the search space is limited to SOS polynomials for multipliers $\mu(\boldsymbol{x})$, while in \eqref{eq:enlargement2}, search space is all polynomial multipliers $\lambda(\boldsymbol{x})$.
\end{Remark}

\begin{Remark}
    By maximizing the constant $\gamma$ in equation \eqref{eq:enlargement2}, we can ensure that the difference between $b^{(k)}(\boldsymbol{x})$ and $b^{(k+1)}(\boldsymbol{x})$ is maximized in the CBC synthesis procedure.
\label{Remark:maximize_gamma}
\end{Remark}

%%%%%%%%%%%%%%%%%%%%%%%%%%%
%%%%%%%%%%%%%%%%%%%%%%%%%%%
%%%%%%%%%%%%%%%%%%%%%%%%%%%
\subsection{Iteratively Constructing and Enlarging CBCs}
\label{sec:Iteratively_Constructing}
With the theoretical foundation in place, we now introduce an iterative algorithm for synthesizing CBCs.  
By combining SOSP \eqref{eq:CBC_synth} together with the conditions of Proposition \ref{proposition1}, and the idea explained in Remark \ref{Remark:maximize_gamma}, we propose the following SOSP formulation as our second result.

\begin{Proposition}
    \label{Proposition2}
    Let system \eqref{eq:affine_sys}, the sets of unsafe states, and input constraints be given as in Theorem \ref{CBC_synth}. Let $b_0(\boldsymbol{x})$ be a CBC for system \eqref{eq:affine_sys}, constructed by SOSP \eqref{eq:CBC_synth}. If   
    \begin{subequations}
    \label{eq:Enlarge_B}
        \begin{equation*}
            \underset{\sigma_{i,j}(\boldsymbol{x}), \mu(\boldsymbol{x})\in\Sigma{[\boldsymbol{x}]}, \ b(\boldsymbol{x}), \boldsymbol{u}(\boldsymbol{x}), \lambda_1(\boldsymbol{x}),\lambda_2(\boldsymbol{x}),\lambda_3(\boldsymbol{x})\in\mathbb{R}{[\boldsymbol{x}]}, \ \gamma > 0}
            {\textnormal{ maximize}} \gamma
        \end{equation*}
        \begin{equation}
        \label{eq:Enlarge_B1}
         \textnormal{s.t.} \ -b(\boldsymbol{x}) + \sum_{j=1}^{m_i}{\sigma_{i,j}(\boldsymbol{x}) s_{i,j}(\boldsymbol{x})} -\epsilon \in \Sigma[\boldsymbol{x}], \quad i = 1, \ldots, n,
        \end{equation}
        \begin{equation}
        \label{eq:Enlarge_B2}
        b(\boldsymbol{x}) - \mu(\boldsymbol{x}) b_0(\boldsymbol{x})\in\Sigma{[\boldsymbol{x}]},
        \end{equation}
        \begin{equation}
        \label{eq:Enlarge_B4}
         A_u \boldsymbol{u}(\boldsymbol{x}) + c_u - \lambda_1(\boldsymbol{x}) b(\boldsymbol{x}) \in \Sigma[\boldsymbol{x}],
        \end{equation}
        \begin{equation}
        \label{eq:Enlarge_B3}
          \frac{\partial b(\boldsymbol{x})}{\partial \boldsymbol{x}}(f(\boldsymbol{x}) + g(\boldsymbol{x}) \boldsymbol{u}(\boldsymbol{x})) - \lambda_2(\boldsymbol{x}) b(\boldsymbol{x}) \in \Sigma[\boldsymbol{x}],
        \end{equation}
        \begin{equation}
        \label{eq:Enlarge_B5}
         b(\boldsymbol{x}) - \lambda_3(\boldsymbol{x})b_0(\boldsymbol{x})-\gamma \in\Sigma{[\boldsymbol{x}]}.
        \end{equation}
    \end{subequations}
is feasible for some $\epsilon$, then $b(\boldsymbol{x})$ is CBC for system \eqref{eq:affine_sys} and, ~~~~~~~~~~~~~~~~~~  $\{\boldsymbol{x} \mid b_0(\boldsymbol{x})\ge0\}\subset\{\boldsymbol{x} \mid b(\boldsymbol{x})\ge0\}$.
\end{Proposition}

\begin{proof}
    The proof follows similarly to the proofs of Theorem \ref{CBC_synth} and Proposition \ref{proposition1}.
\end{proof}
\vspace{-0.3 cm}
In \eqref{eq:Enlarge_B}, we solve a maximization problem to synthesize an enlarged safe set corresponding to $b(\boldsymbol{x})$ with respect to previously constructed safe set by $b_0(\boldsymbol{x})$. Next, we explain how we address bilinearity and solve SOSP \eqref{eq:Enlarge_B}.

%%%%%%%%%%%%%%%%%%%%%%%%%%%
\subsubsection{Initialization}
First, we define $\mathcal{X}_0$ as a small set around a point of interest (e.x., the stable equilibrium of the system) by assuming $b_0(\boldsymbol{x})=\delta-V_0(\boldsymbol{x})$, where $V_0(\boldsymbol{x})$ is quadratic cost-to-go that can be calculated by solving Linear Quadratic Regulator (LQR) problem around the point of interest and $\delta$ is a small positive real number.
We then solve SOSP \eqref{eq:CBC_synth} by setting $b(\boldsymbol{x}) = b_0(\boldsymbol{x})$ to determine the required multipliers and initialize $u(\boldsymbol{x})$.
Notice that \eqref{eq:CBC_synth} is not bilinear with known $b(\boldsymbol{x})$.
%%%%%%%%%%%%%%%%%%%%%%%%%%%
\subsubsection{Finding Enlarged $\mathcal{S}$} 
With $\lambda_1(\boldsymbol{x})$, $\lambda_2(\boldsymbol{x})$, and $u(\boldsymbol{x})$ initialized, all bilinear terms in SOSP \eqref{eq:Enlarge_B} become linear. Consequently, the problem is convex and can be efficiently solved to find $b(\boldsymbol{x})$ and maximize $\gamma$.

%%%%%%%%%%%%%%%%%%%%%%%%%%%
\subsubsection{Refining $\boldsymbol{u}(\boldsymbol{x})$ and Multipliers $\lambda_1(\boldsymbol{x})$ and $\lambda_2(\boldsymbol{x})$}
Once the enlarged $\mathcal{S}$ is obtained by solving SOSP \eqref{eq:Enlarge_B}, we refine the synthesized controller and compute new polynomial multipliers for the next iteration by solving the following SOSP:
\vspace{-0.3 cm}
\begin{subequations}
    \label{eq:Refine_u}
    \begin{equation*}
        \textnormal{Find }  \boldsymbol{u}(\boldsymbol{x}),\lambda_1(\boldsymbol{x}),\lambda_2(\boldsymbol{x})\in\mathbb{R}{[\boldsymbol{x}]}
    \end{equation*}
    \vspace{-0.3 cm}
    \begin{equation}
        \textnormal{s.t.}\quad A_u\boldsymbol{u}(\boldsymbol{x})+c_u-\lambda_{1}(\boldsymbol{x})b(\boldsymbol{x})
         \in\Sigma{[\boldsymbol{x}]},\quad\ 
    \end{equation}
    \vspace{-0.3 cm}
    \begin{equation}
         \frac{\partial b(\boldsymbol{x})}{\partial \boldsymbol{x}}(f(\boldsymbol{x}) + g(\boldsymbol{x}) \boldsymbol{u}(\boldsymbol{x}))-\lambda_{2}(\boldsymbol{x})b(\boldsymbol{x})\in\Sigma{[\boldsymbol{x}]}.
    \end{equation}
\end{subequations}
The SOSP \eqref{eq:Refine_u} is equivalent to \eqref{eq:CBC_synth} without constraints \eqref{eq:theo_a} and \eqref{eq:theo_a2}, since the fixed $b(\boldsymbol{x})$ always satisfies \eqref{eq:theo_a} and \eqref{eq:theo_a2}.

%%%%%%%%%%%%%%%%%%%%%%%%%%%
\subsubsection{Termination of Iterative Procedure}
The bilinear alternation process continues until the acquired $\gamma$ falls below a predefined threshold. This indicates that the enlargement is sufficiently small, allowing the termination of the algorithm.

Algorithm \ref{Alg1} summarizes our approach for constructing a CBC for a given dynamical system and unsafe set.
% \vspace{-0.25 cm}
\subsection{Using The Synthesized CBC to Generate Safe Control}
Having found a CBC function $b(\boldsymbol{x})$, and a given nominal control action $\boldsymbol{u}_N$, we can calculate a safe action, by solving the following QP, 
% Although Algorithm \ref{Alg1} constructs a controller $\boldsymbol{u}(\boldsymbol{x})$ at each iteration, we can alternatively compute a min-norm control signal for system \eqref{eq:affine_sys} with a nominal control $\boldsymbol{u}_N$ by solving  
\vspace{-0.3 cm}
\begin{subequations}
\label{eq:QP_cost}
    \begin{equation}
        \underset{\boldsymbol{u}\in\mathcal{U}}
        {\textnormal{minimize }} |\boldsymbol{u}-\boldsymbol{u}_N|
    \end{equation}
    \vspace{-0.3 cm}
    \begin{equation}
    \label{eq:QP_constraint}
     \textnormal{s.t.} \quad \frac{\partial b(\boldsymbol{x})}{\partial \boldsymbol{x}}(f(\boldsymbol{x}) + g(\boldsymbol{x}) \boldsymbol{u}) \ge -\eta\,b(\boldsymbol{x})
    \end{equation}
\end{subequations}
for a sufficiently large $\eta\in\mathbb{R}^{+}$ to relax the constraint inside the safe set and enforce it on the boundary, when $b(\boldsymbol{x})=0$.

\begin{algorithm}
    \caption{AFTC for AFS}
    \begin{algorithmic}[1]  
        \State Start with $b_0(\boldsymbol{x})$, $k := 0$, and \texttt{converged} := False.
        \State \textbf{Initialization:} Given $b^{(k)}(\boldsymbol{x})=b_0(\boldsymbol{x})$, find $\boldsymbol{u}^{(k)}(\boldsymbol{x})$ and $\lambda_1^{(k)}(\boldsymbol{x}),\ \lambda_2^{(k)}(\boldsymbol{x})$ by solving \eqref{eq:CBC_synth}.

        \While{not \texttt{converged}}
            \State $k \mathrel{+}= 1$
            \State Given $\boldsymbol{u}^{(k-1)}(\boldsymbol{x})$ and $\lambda_1^{(k-1)}(\boldsymbol{x}),\ \lambda_2^{(k-1)}(\boldsymbol{x})$, solve SOSP \eqref{eq:Enlarge_B} to acquire $b^{(k)}(\boldsymbol{x})$ and $\gamma^{(k)}$.
            \If {$\gamma^{(k)} \leq \text{Threshold}$}
                \State \texttt{converged} := True
            \Else
                \State Given $b^{(k)}(\boldsymbol{x})$, solve SOSP \eqref{eq:Refine_u} to acquire $\boldsymbol{u}^{(k)}(\boldsymbol{x})$ and $\lambda_1^{(k)}(\boldsymbol{x}),\ \lambda_2^{(k)}(\boldsymbol{x})$. Then, set
                $b_0(\boldsymbol{x})=b^{(k)}(\boldsymbol{x})$.
            \EndIf
        \EndWhile
    \end{algorithmic}
    \label{Alg1}
\end{algorithm}  
%%%%%%%%%%%%%%%%%%%%%
%%%%%%%%%%%%%%%%%%%%%
%%%%%%%%%%%%%%%%%%%%%
\section{Simulation Results}
\label{Results}
We evaluate the proposed approach using two numerical examples. 
First, we synthesize a CBC for the well-known 2D benchmark for nonlinear systems, Van der Pol oscillator, and compare the size of the safe set generated by our method with two state-of-the-art methods. 
Subsequently, we apply our method to synthesize a CBC for a multiple-input 3D dynamical system adapted from \cite{wang2018permissive}.
All simulations are conducted in MATLAB on a computer with an Intel Core i5 CPU. We utilize SOSTOOLS \cite{sostools} to formulate SOSPs and solve them using Mosek \cite{mosek}. All codes are available at GitHub\footnote{\url{https://github.com/Naeim-Eb/CBC-SOS}}.

% We also provide simulations of trajectories for both systems, starting on the boundary of constructed the safe set by our method. To simulate trajectories, we use a CBC-QP controller as \eqref{eq:QP_cost} with $\boldsymbol{u}_N$ defined as a Proportional-Derivative (PD) controller which aims to stabilize system to the origin.% to simulate the trajectories.

We provide trajectory simulations for both systems, starting from the boundary of the safe set constructed by our method. To generate these trajectories, we use a CBC-QP controller as in \eqref{eq:QP_cost}, with $\boldsymbol{u}_N$ defined as a Proportional-Derivative (PD) controller aiming to stabilize the system to the origin.

% We also provide simulations of trajectories for both systems, starting within the safe set constructed by our CBC. To simulate the trajectories, we assume a potentially unsafe feedback controller as the nominal control $\boldsymbol{u}_N$, which is a Prpporional-Derivative (PD) controller that aims to stabilize the system to the origin and solve a QP that minimally modifies the nominal control input to satisfy the CBC condition \eqref{eq:BC_cond_3} as a constraint. To ensure that the constraint activates only near the boundary of the safe set, we modify \eqref{eq:BC_cond_3} as follows:
% \begin{equation}
%     \frac{\partial b(\boldsymbol{x})}{\partial \boldsymbol{x}}(f(\boldsymbol{x}) + g(\boldsymbol{x}) \boldsymbol{u}) \ge -\eta\,b(\boldsymbol{x})
% \label{eq:CBC_QP_constraint}
% \end{equation}
% where $\eta$ is a positive scalar. 
% The primary advantage of using CBC in constraint \eqref{eq:CBC_QP_constraint}, as opposed to a CBF, is its independence from $\eta$, since $\eta$ is not involved in the synthesis procedure. Consequently, our CBC-based approach allows for adjusting conservativeness during runtime without being restricted by a fixed $\eta$ determined during synthesis.

\subsection{Example 1}
We consider the Van der Pol oscillator system defined as:
\begin{equation*}
\label{eq:example1}
    \begin{bmatrix} \dot{x}_1\\\dot{x}_2\end{bmatrix} =
    \begin{bmatrix} x_2\\(1-x_1^2)x_2 - x_1\end{bmatrix} +
    \begin{bmatrix} 0\\1 \end{bmatrix}u,
\end{equation*}
with the input constraint $|u|\leq 1$ and the following unsafe state regions:
\begin{equation*}
    \mathcal{X}_{u_1}=\{\boldsymbol{x} \mid 4-x_1^2<0\}, \quad \mathcal{X}_{u_2}=\{\boldsymbol{x} \mid 4-x_2^2<0\},
\end{equation*}
\begin{equation*}
    \mathcal{X}_{u_3}=\{\boldsymbol{x} \mid (x_1-1)^2+(x_2-1)^2-0.04<0\},
\end{equation*}
\begin{equation*}
    \mathcal{X}_{u_4}=\{\boldsymbol{x} \mid (x_1+1)^2+(x_2+1)^2-0.04<0\},
\end{equation*}
\begin{equation*}
    \mathcal{X}_{u_5}=\{\boldsymbol{x} \mid (x_1+1)^2+(x_2-1)^2-0.04<0\}.
\end{equation*}

\begin{figure}
    \centering
    \includegraphics[width=\columnwidth, trim= 0.3cm 1.65cm 20.6cm 0.25cm, clip]{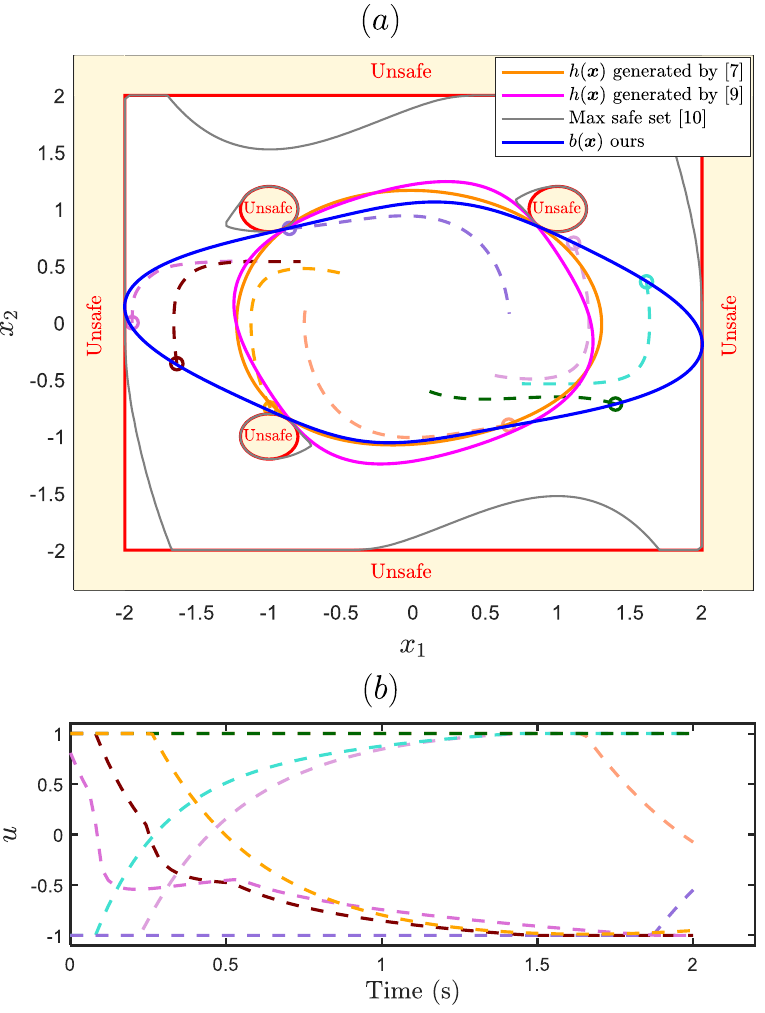}
    \caption{(a) Synthesized safe set for Example 1 using our method vs. \cite{dai2023convex, zhao2023convex}, along with the maximum safe set generated by Hamilton-Jacobi backward reachable sets \cite{bansal2017hamilton}. The figure also shows safe control generated via our CBC-QP framework, guiding random trajectories from the boundary back inside.  
    (b) Control inputs for the trajectories, all within input bounds.
}
    \label{fig:VanDerPole_graph}
    \vspace{-0.3 cm}
\end{figure}

We apply Algorithm \ref{Alg1} with $b(\boldsymbol{x})$ and $u(\boldsymbol{x})$ as polynomials of degree 4 and 3, respectively, and all multipliers $\lambda_i(\boldsymbol{x})$ and $\sigma_i(\boldsymbol{x})$ of degree 4. The resulting safe set is shown in Fig.~\ref{fig:VanDerPole_graph}(a).

Using our CBC, we generate safe control signals via \eqref{eq:QP_constraint} and simulate eight trajectories starting from different points on the safe set boundary, as depicted in Fig.~\ref{fig:VanDerPole_graph}(a). The corresponding control inputs, shown in Fig.~\ref{fig:VanDerPole_graph}(b), satisfy input constraints.

To compare our method with two state-of-the-art approaches \cite{zhao2023convex, dai2023convex} for input-constrained CBF synthesis, we generated CBFs using these methods and presented the safe sets in Fig.~\ref{fig:VanDerPole_graph}. Additionally, we included the maximum safe set computed via Hamilton-Jacobi Backward Reachable Sets (HJ-BRS) \cite{bansal2017hamilton}. For a fair comparison, we used similar polynomial degrees for the CBFs and multipliers. However, the method in \cite{dai2023convex} required increasing the multiplier degree to 20 to achieve its maximum capacity.

As it can be seen, the safe set generated with our method covers in most part a larger portion of the state space, demonstrating less conservatism. The synthesis times were 29.36s for \cite{zhao2023convex}, 231.19s for \cite{dai2023convex}, and 18.44s for our method. The synthesized polynomials are provided in Appendix A.

%Figure~\ref{fig:VanDerPole_graph} illustrates that the safe set from our method covers in most part a larger portion of the state space, demonstrating less conservatism in Example 1. The synthesis times were 29.36s for \cite{zhao2023convex}, 231.19s for \cite{dai2023convex}, and 18.44s for our method. The synthesized polynomials are provided in Appendix A.

\subsection{Example 2}
\begin{figure}
\centering
\includegraphics[width=\columnwidth, trim= 0.2cm 4.91cm 11.6cm 0.2cm, clip]{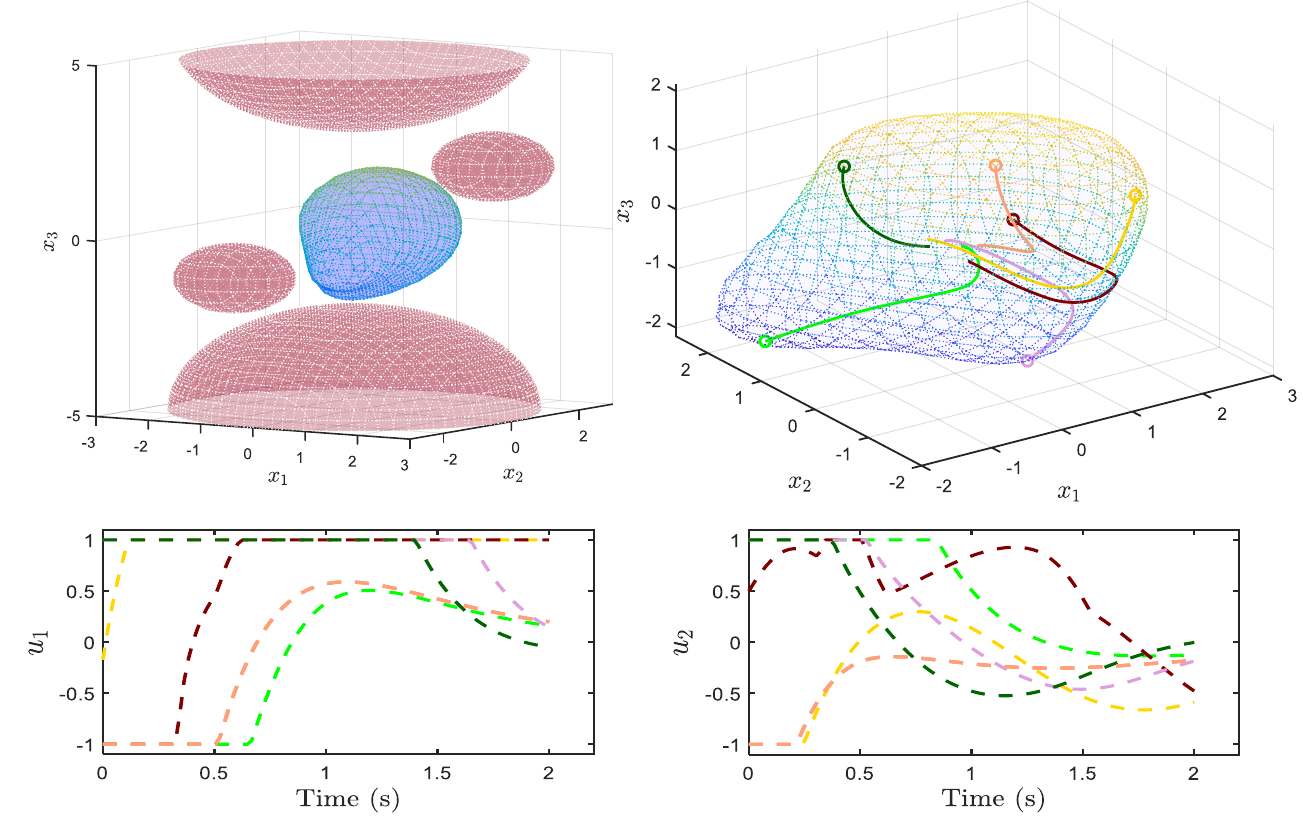}
\caption{Synthesized safe set for Example 2 (blue) with unsafe zones (red) (top left).  
Simulated trajectories from random boundary points (top right).  
Control inputs for each trajectory, all within bounds (bottom).
}
\label{fig:3D}
\vspace{-0.3 cm}
\end{figure}

As a second example, and to validate the efficacy of our method for more complex systems, we synthesize a CBC for the system described in Example 2 of \cite{wang2018permissive}. The system's state vector is defined as $\boldsymbol{x} = [x_1, x_2, x_3]^T \in \mathbb{R}^3$, and the control input vector is defined as $\boldsymbol{u} = [u_1, u_2]^T \in \mathbb{R}^2$. We consider similar unsafe regions. %, shown in red in Fig.~\ref{fig:3D}
%defined as $\mathcal{X}_{u_i} = \{\boldsymbol{x} \mid q_i(\boldsymbol{x}) < 0\}$ for $i = 1, \ldots, 4$. 
In addition to state constraints, we impose input constraints as $|u_j| \le 1$ for $j = 1, 2$. The safe set generated using Algorithm~\ref{Alg1} is depicted in Fig.~\ref{fig:3D} (top left).

% \begin{figure}
% \centering
% %\includegraphics[width=7 cm]{Graphs/3D_CBC2.pdf}
% \includegraphics[width=0.75\columnwidth, trim= 0.2cm 8.1cm 20.5cm 0.2cm, clip]{Graphs/CBC_3D.pdf}
% \caption{Synthesized safe safe set with our method for the system in Example 2, shown in blue. Red regions denote unsafe zones.}
% \label{fig:3D}
% \end{figure}

% To verify the validity of the proposed CBC, we simulate 6 trajectories starting from different points along the boundary of the safe set, as shown in Fig.~\ref{fig:3D} (top right). As observed, the controller ensures that all trajectories remain within the synthesized set while satisfying the input constraints as depicted in Fig.~\ref{fig:3D} (bottom graphs), thereby confirming the controlled invariance of the synthesized safe set.
To validate the proposed CBC, we simulate six trajectories starting from different points on the safe set boundary, as shown in Fig.~\ref{fig:3D} (top right). The controller keeps all trajectories within the synthesized set while satisfying input constraints, as shown in Fig.~\ref{fig:3D} (bottom graphs), confirming the controlled invariance of the safe set.

% \begin{figure}
%     \centering
%     \includegraphics[width=\columnwidth, trim= 0.2cm 9cm 11.6cm 0.2cm, clip]{Graphs/trajectories_3D_ver2.pdf}
    % \caption{Simulated trajectories from various random points on the boundary of the synthesized safe set with our method for Example 2 (left). 
    % control input corresponding to the various trajectories in the left graph for each control, all satisfying input bounds.
    % }
%     \label{fig:trajectories_3D}
% \end{figure}

%%%%%%%%%%%%%%%%%%%%%
%%%%%%%%%%%%%%%%%%%%%
%%%%%%%%%%%%%%%%%%%%%

%%%%%%%%%%%%%%%%%%%%%
%%%%%%%%%%%%%%%%%%%%%
%%%%%%%%%%%%%%%%%%%%%
% \textcolor{red}{\section{Conclusion}}
% \label{Conclusion}
% This paper presents an iterative method for designing controlled-invariant (safe) sets for dynamical systems with predefined unsafe states. To achieve this, we formulate an SOSP to synthesize CBCs starting with a given set of initial conditions. We then propose conditions that guarantee the progressive enlargement of the safe set in an iterative process and provide theoretical proof of this enlargement.  
% Finally, we validate the effectiveness of our method through numerical simulations of 2D and 3D dynamical systems with single- and multi-input configurations. In a 2D example, we compare our synthesized safe set with two state-of-the-art methods and show that our approach produces larger controlled-invariant (safe) sets.  
\section{Conclusion}
\label{Conclusion}

This paper introduced a method for synthesizing controlled-invariant sets for nonlinear polynomial control-affine systems using Control Barrier Certificates (CBCs). We formulated CBC conditions as Sum-of-Squares (SOS) constraints and solved them via an SOS Program (SOSP). To handle complex unsafe state representations, we generalized existing SOSPs for CBC synthesis and proposed an iterative algorithm that progressively enlarges the safe set constructed by the synthesized CBCs. Our theoretical analysis guarantees strict safe set enlargement at each iteration.  
We validated our approach through numerical simulations in 2D and 3D for single-input and multi-input systems. Empirical results demonstrate that the safe set generated by our method covers in most part a larger portion of the state space compared to two state-of-the-art techniques.

Our future work includes extending the proposed method to synthesize CBCs for general non polynomial systems.

%%%%%%%%%%%%%%%%%%%%%%
%%%%%%%%%%%%%%%%%%%%%%
%%%%%%%%%%%%%%%%%%%%%%

\bibliographystyle{ieeetr}
\bibliography{mybibfile}

% \appendices
\section{Appendix A}
\label{app:FirstAppendix}
The synthesized functions that have been used in Fig.~\ref{fig:VanDerPole_graph} (a) are as follows.

CBF synthesized by method in \cite{zhao2023convex}:
\begin{equation*}
    \begin{aligned}
        h(\boldsymbol{x}) = & -0.031 x_1^4 - 0.032 x_1^3 x_2 + 0.004 x_1^2 x_2^2 \\
        & + 0.007 x_1 x_2^3 - 0.002 x_2^4 + 0.065 x_1^3 \\
        & - 0.080 x_1^2 x_2 - 0.032 x_1 x_2^2 - 0.035 x_2^3 \\
        & - 0.580 x_1^2 + 0.048 x_1 x_2 - 0.798 x_2^2 \\
        & - 0.051 x_1 + 0.121 x_2 + 1,
    \end{aligned}
\end{equation*}
CBF synthesized by method in \cite{dai2023convex}:
\begin{equation*}
    \begin{aligned}
        h(\boldsymbol{x}) = & -0.708 x_1^4 - 1.481 x_1^2 x_2^2 + 0.863 x_1 x_2^3 \\
        & - 0.648 x_2^4 - 0.000 x_1^2 x_2 + 0.408 x_1^2 \\
        & - 0.616 x_1 x_2 + 0.286 x_2^2 + 1,
    \end{aligned}
\end{equation*}
our CBC:
\begin{equation*}
    \begin{aligned}
        b(\boldsymbol{x}) = & -95.709 x_1^4 - 105.270 x_1^3 x_2 - 653.887 x_1^2 x_2^2 \\ 
        & + 175.891 x_1 x_2^3 - 229.534 x_2^4 - 20.112 x_1^3 \\
        & - 59.575 x_1^2 x_2 - 52.366 x_1 x_2^2 - 59.885 x_2^3 \\ & + 258.824 x_1^2 - 48.561 x_1 x_2 - 127.349 x_2^2 \\
        & + 68.246 x_1 + 71.585 x_2 + 419.753.
    \end{aligned}
\end{equation*}

%\section{\textcolor{red}{Some notes to be handled}}
% \begin{figure}[H]
% \centering
% %\includegraphics[width=7 cm]{Graphs/3D_CBC2.pdf}
% \includegraphics[width=0.75\columnwidth, trim= 0.2cm 4.87cm 20cm 0.2cm, clip]{Graphs/HJ_BRT.pdf}
% \caption{ Hamilton Jacobi Backward Reachable Tube}
% \label{fig:HJ_BRT}
% \end{figure}

% \begin{enumerate}
%     \item I have not define class-$\mathcal{K}$ function because it does not have any relation with CBC. Also, I have not used that throughout my methodology.
% \end{enumerate}

\end{document}